\documentclass{revtex4}
\usepackage{psfig}
\usepackage{subfigure}
\usepackage{graphicx}
\usepackage{doublespace}
\usepackage{amsmath}

\begin{document}

\doublespace

\begin{abstract}
We consider synchronization of coupled chaotic systems and propose an adaptive strategy that aims at evolving the strength of the coupling to achieve stability of the synchronized evolution.
We test this idea in a simple configuration in which two chaotic systems are unidirectionally coupled (a sender and a receiver) and we study conditions for the receiver to adaptively synchronize with the sender.
Numerical simulations show that, under certain conditions, our strategy is successful in dynamically evolving the coupling strength until it converges to a value that is compatible with synchronization.
\end{abstract}


\author{Francesco Sorrentino}
\affiliation{Universit{\`a} degli Studi di Napoli Parthenope, 80143 Napoli, Italy. \\ Institute for Research in Electronics and Applied Physics, University of Maryland, College Park, Maryland 20742.}

\title{Adaptive coupling for achieving stable synchronization of chaos}
\maketitle

\section{Introduction}

It is known that stability of the synchronized solution of coupled dynamical systems depends on the strength of the interaction.
Motivated by the observation that real-world networks are characterized by evolving adapting connections, 
some studies have started to take in consideration how the time variability of the strength of the couplings may affect the synchronization dynamics.
A few papers have appeared on adaptive strategies, in which the coupling strengths are dynamically adjusted based on information on the states of the systems,  e.g., in order to enhance or guarantee synchronization.
In this paper we address the problem of how to evolve the coupling strength from a given initial condition to make it converge to a value which corresponds to stability of the synchronized evolution.

 Previous works have addressed adaptive synchronization of chaos. For instance, in \cite{SOTT,SOTT2},  a problem was studied in which adaptation was needed in order to achieve synchronization in the presence of external unpredictable events affecting the communication between the coupled systems. As an applicative example, consider the case of two sensors that seek to synchronize through a continuous signal that they exchange via wireless communication;  if an object moves across the communication pathway between the sensors, this causes an attenuation on the strength of the received signal, which may disrupt synchronization.  
 Therefore, in \cite{SOTT,SOTT2}, adaptive strategies were introduced, having the aim of maintaining synchronization with respect to such  external (unpredictable) perturbations. 

{Here, we will focus on a different but related problem, namely we consider a simple configuration in which a sender is connected to a receiver and communication between the two is such that attenuations affecting the received signal are negligible; the problem is rather for the receiver to adaptively choose an appropriate coupling form (in what follows we will specify our proposed problem to be that of choosing an appropriate coupling strength) with the received signal 
  in order to achieve synchronization with the sender. }

%
%
%
%
%
%

To better illustrate this problem, we introduce the equations for the sender and the receiver,
\begin{subequations}\label{masl}
\begin{align}
\dot{x}_1(t)=F(x_1(t)), \label{ma} \\
\dot{x}_2(t)=F(x_2(t))+\gamma(t)(H(x_1(t))-H(x_2(t))), \label{sl}
\end{align}
\end{subequations}
where $x_1(t)$ ($x_2(t)$) is the $m$-dimensional state of the sender (receiver) system, $x_i=[x_{i1},x_{i2},...,x_{im}]$;  $F(x)$  is the dynamics of an uncoupled system (hereafter assumed chaotic), $F:R^m \rightarrow R^m$; $H(x)$ is an output function, $H:R^m \rightarrow R^m$; and $\gamma(t)$ is a time varying scalar function measuring the strength of the coupling. Here we assume that $H(x)$ can be rewritten as $H(x)=\mathcal{H} h(x)$, where $\mathcal{H}=[\mathcal{H}_1,\mathcal{H}_2,...,\mathcal{H}_m]$ is a constant $m$-vector and $h:R^m \rightarrow R$ is a scalar output function. In many practical situations, communication between the connected systems involves only a subset of the dynamical state variables of the systems. Therefore, in this paper we consider a situation in which synchronization has to be achieved based solely on the scalar signal, $h(x_1(t))$, that system 2 receives from system 1. Our proposed problem is to devise an adaptive strategy to dynamically evolve the coupling strength $\gamma(t)$ from an arbitrary initial condition $\gamma(0)$ to a value that is compatible with synchronization.

{As an example of an application for our proposed strategy (to be specified in Sec. II), we consider a \emph{sender-repeater} communication scenario, in which the first device, say a wifi antenna, sends a signal, and the second identical device needs to reproduce precisely the same signal to other systems that cannot get the original one. 
As another example,  our strategy could be used in applications  that take advantage of synchronization of chaos for identification and prediction of the dynamics of unknown real systems \cite{Abarbanel,Abarbanel2,Abarbanel4,IDTOUT}. In particular, for the problems addressed in \cite{Abarbanel,Abarbanel2,Abarbanel4,IDTOUT}, the receiver is a model system which is meant to replicate the dynamics of an unknown true system (the sender) and a technique is proposed that aims at identifying the true system parameters by 
making the receiver synchronize with the sender. The technique relies on the choice of a coupling strength which is compatible with synchronization, but for this case, an extra difficulty is due to the fact that the dynamical form of the sender's $F(x)$ is unknown at the receiver and therefore one is unable  to compute  the $\gamma$-range of stability (i.e., compute the master stability function; for more details on this subject the reader is referred to the background section) beforehand. Therefore, if one has to deal with such a problem and unless one wants to exclusively rely on a try-and-error approach,   
 it becomes necessary  to introduce an adaptive strategy to dynamically evolve the coupling strength. 
We anticipate that the adaptive strategy that we present here does not use information from the master stability function and is based exclusively on minimization of the squared synchronization error $[h(x_1(t))-h(x_2(t))]^2$. Therefore, we envision our strategy would be particularly useful in applications that use synchronization as a tool to achieve 
parameter estimation and prediction 
of unknown systems (e.g., for all the applications described in \cite{Abarbanel,Abarbanel2,Abarbanel4,IDTOUT}). 

\subsection{Background}

We note that Eqs. (\ref{ma}) and (\ref{sl}) admit a synchronous solution of the type,
\begin{equation}
x_2(t)=x_1(t), \label{s}
\end{equation}
which obeys the sender evolution (\ref{ma}). Our proposed problem is to devise a strategy to evolve $\gamma(t)$ from a given initial condition $\gamma(0)$ in order for the receiver system to synchronize with the sender.
Let us now assume that $\gamma(t)$ is constant and equal to $\gamma$.
Following \cite{FujiYama83,Afraim,Pe:Ca,Yang1}, we know that depending on the choice of the functions $F$ and $H$, synchronization is possible in a certain range of values of $\gamma$, say $\Gamma$ .
To see this, let us linearize the receiver system equation (\ref{sl}) about (\ref{s}), obtaining,
\begin{equation}
\delta \dot{x}_2(t)=[DF(x_1(t)) - \gamma DH(x_1(t))] \delta x_2(t), \label{lin}
\end{equation}
where $DF(x_1(t))$ and $DH(x_1(t))$ are the Jacobians of the functions $F$ and $H$ evaluated about the synchronous solution (\ref{s}).
Then stability of the synchronous solution depends on the maximum Lyapunov exponent associated with Eq. (\ref{lin}) (this is a classic result, see \cite{FujiYama83,Afraim,Pe:Ca,Yang1} for more details). Once the functions $F$ and $H$ are given, it becomes possible to study the dependence of the maximum Lyapunov exponent of (\ref{lin}) on $\gamma$. The function that associates $\gamma$ with the maximum Lyapunov exponent of (\ref{lin}) is usually called a master stability function \cite{Pe:Ca}.

From \cite{FujiYama83,Afraim,Pe:Ca,Yang1} and subsequent papers, we know that the range of values of $\gamma$, say $\Gamma$, in which the master stability function associated with Eq. (\ref{lin}) is negative, can be either bounded or unbounded.  An example of a bounded range for $\gamma$ is reported in Fig. 1, where for $F(x)$ being the equation of the R\"ossler system, $m=3$, and $h(x_i)=x_{i1}$, $\mathcal{H}=[1,0,0]$, the master stability function is found to be negative in the range $\Gamma=[0.3,5]$. Moreover, we observe from Fig. 1 that the master stability function has a minimum at $\gamma \simeq 1.7$ (corresponding to the maximum rate of contraction for (\ref{lin}) and therefore the fastest possible convergence to the synchronized evolution). Another example of a bounded stability range $\Gamma$ is reported in \cite{Yang1} for $F(x)$ being the equation of the Lorenz system, $m=3$, and $h(x_i)=x_{i3}$, $\mathcal{H}=[0,0,1]$.

\begin{figure}[h]
\centerline{\psfig{figure=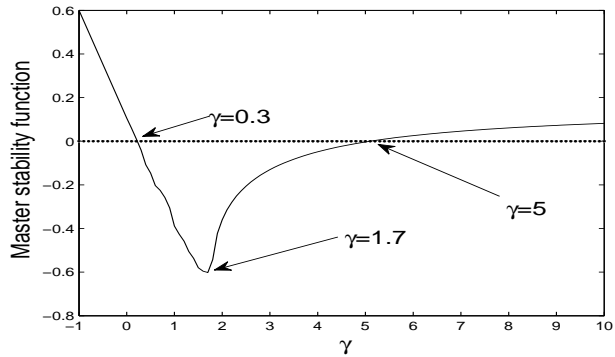,width=9cm,height=5cm}}
\caption{\small Master stability function corresponding to Eq. 
 (\ref{lin}), for our choices of $F(x)$ being the Rossler equation (\ref{F}) and $h(x_i)=x_{i1}$, $\mathcal{H}=[1,0,0]$. The zero ordinate line is shown as a guide to the eye (dotted line). \label{A1}}
\end{figure}

 The problem we address in this paper (a strategy to evolve the strength of the coupling for achieving stable synchronization) has already been addressed in a number of papers \cite{Zh:Ku06,Delellis1,Delellis2, Abarbanel}. Namely, in \cite{Zh:Ku06,Delellis1,Delellis2}, the following adaptive strategies were proposed,
\begin{subequations}
\begin{align}
\dot{\gamma}(t)=\alpha |h(x_1(t))-h(x_2(t))|, \quad \mbox{ in [12,13]},
\label{str} \\
\dot{\gamma}(t)=\alpha \frac{|h(x_1(t))-h(x_2(t))|}{1+|h(x_1(t))-h(x_2(t))|}, \quad \mbox{ in [11]},
\label{str2}
\end{align}
\end{subequations}
where $\alpha>0$ is a suitable scalar gain. Equation  (\ref{str2}) is approximately equal to Eq. (\ref{str}) for $|h(x_1(t))-h(x_2(t))| \ll 1$, and is approximately equal to $\dot{\gamma}(t)=\alpha$, for $|h(x_1(t))-h(x_2(t))| \gg 1$.
It is worth noting  that both strategies are based on increasing the coupling strength $\gamma$ proportionally with the absolute value of the synchronization error (with a saturation in the case of (\ref{str2})).

We note that  the strategies proposed in \cite{Zh:Ku06,Delellis1,Delellis2} may not be the best suited to deal with coupled dynamical systems whose stability is governed by a master stability function as that shown in Fig. 1.
 For example, for the case shown in Fig. 1, if $\gamma(0)$ is larger than $5$, then increasing $\gamma$ will not be effective in order to achieve synchronization. Moreover, even if $\gamma(0) \in \Gamma$, strategies (\ref{str},\ref{str2}) may eventually increase $\gamma$  outside the range $\Gamma$ 
 (e.g., beyond $5$ in Fig. 1) and yield desynchronization even in a case in which the systems would have synchronized with no adaptation (i.e., with $\dot \gamma=0$).

Another adaptive strategy has been proposed in \cite{Abarbanel}, based on the following adaptation for $\gamma(t)$,
\begin{equation}
\dot{\gamma}(t)=-a \gamma(t) + g [(h(x_1(t))-h(x_2(t)))^2], \label{AB}
\end{equation}
where $a>0$   and $g(x)$ is a function such that $g(x) \approx x$ for small $x$ and is bounded by a constant $C$ for large $x$.  We note 
from (\ref{AB}), that for large $t$, $\gamma(t)$ converges to $a^{-1} g [(h(x_1(t))-h(x_2(t)))^2]$ and therefore for a monotonely increasing $g(x)$, $\gamma(t)$ increases monotonely with the squared synchronization error. 
Hence, (\ref{AB}) is subject to the same sort of limitations as (\ref{str},\ref{str2}) when stability of the synchronized evolution is described by a master stability function as that shown in Fig. 1.  

In what follows, we will present a novel adaptive strategy, which is aimed at evolving $\gamma(t)$ from an arbitrary initial condition $\gamma(0)$ to converge on $\Gamma$.
We note that the following limitations may apply to some of the previously reported strategies (\ref{str},\ref{str2},\ref{AB}),
\begin{enumerate}
  \item 
  The strategies (\ref{str},\ref{str2},\ref{AB}) are based on increasing the coupling strength with the synchronization error. Yet, for the case of a bounded range of stability $\Gamma$,   either increasing or \emph{decreasing} the coupling strength $\gamma$ may be needed in order to achieve synchronization.
  \item The strategies (\ref{str},\ref{str2},\ref{AB}) are dependent on the choice of the initial condition $\gamma(0)$.
   \item The strategies (\ref{str},\ref{str2},\ref{AB}) do not seek to make the coupling $\gamma(t)$ converge in a neighborhood of its optimal value (e.g., the minimum of the master stability function which is observed to be at about $\gamma=1.7$ in Fig. \ref{A1}).
\end{enumerate}

In what follows,  we focus on the common situation that the $\gamma$-range of stability is 
bounded from both below and above (e.g., as shown for the case of the R\"ossler systems in Fig. \ref{A1}). We devise an adaptive strategy that is based on both increasing/decreasing $\gamma(t)$ in order to reach synchronization and will be shown to be independent of the initial condition $\gamma(0)$, as far as $\gamma(0)$ is not too distant from $\Gamma$. Moreover, though our problem is characterized by a continuous of solutions (represented by the range $\Gamma$), our hope is that  $\gamma(t)$ will converge not too far away from the value of $\gamma=1.7$ which corresponds to the minimum of the master stability function, that is to the maximum rate of contraction towards the synchronization manifold. 

We wish to emphasize that our strategy described here is only a possible alternative approach to others already presented in the literature (e.g., \cite{Zh:Ku06,Delellis1,Delellis2,Abarbanel}) and we assume that many other possible solutions can be found to the general problem addressed in this paper. At the same time, we hope our attempt will motivate further studies in the same direction, which will provide better and better solutions to our proposed problem.

The rest of the paper is organized as follows. In Sec. II, we present our adaptive strategy. In Sec. III, we show the results of numerical simulations involving our adaptive strategy. In Sec. IV, the conclusions are presented.

%

\section{Adaptive strategy}

In this section we present our adaptive strategy. We introduce a potential/cost function $\Psi(t)$,
\begin{equation}
\Psi(t)=<[h(x_1(t))-h(x_2(t))]^2>_{\nu}, \label{psi}
\end{equation}
where $<G(t)>_{\nu}$ denotes the sliding exponential average $\int^t e^{-\nu (t-t')} G(t') dt'$. From (\ref{psi}), we note that $\Psi(t)\geq 0$; moreover, 
due to the chaotic nature of the $x$'s, $\Psi(t)=0$ can only be realized if $x_1(t)$ and $x_2(t)$ are in the synchronization manifold, i.e., $x_1(t)=x_2(t)$. Thus we seek to evolve $\gamma(t)$ in order to minimize $\Psi(t)$. To this aim we introduce the following gradient descent relation,
\begin{equation}
\frac{d \gamma(t)}{dt}= -\beta \frac{d \Psi}{d \gamma}= 2 \beta <[h(x_1(t))-h(x_2(t))]\frac{\partial{h}}{\partial x_2}^T  \frac{d x_2}{d \gamma}>_{\nu}= 2 \beta G , \label{gd}
\end{equation}
where $\beta>0$,  and
\begin{equation}
\frac{d{G}(t)}{dt}=-\nu G(t) +[h(x_1(t))-h(x_2(t))]\frac{\partial{h}}{\partial x_2}^T \frac{d x_2}{d \gamma}.
\end{equation}

We are interested in how $\frac{d x_2}{d \gamma}$ evolves in time. To this aim, we write,
\begin{equation}
\frac{d}{dt}\frac{d x_2}{d \gamma}=\frac{d \dot{x}_2}{d \gamma}=[\frac{\partial F}{\partial x_2} -\gamma \frac{\partial{H}}{\partial x_2}] \frac{d x_2}{d \gamma}+[H(x_1(t))-H(x_2(t))].
\end{equation}
To conclude, our adaptive strategy is fully described by the following set of differential equations,
\begin{subequations}\label{GP}
\begin{align}
\dot{\gamma}=2 \beta G,   \label{gp} \\
\dot{G}=-\nu G(t) +[h(x_1(t))-h(x_2(t))]\frac{\partial{h}}{\partial x_2}^T  \xi, \label{gp2}  \\
\dot{\xi}=[\frac{\partial F}{\partial x_2} -\gamma \frac{\partial{H}}{\partial x_2}] \xi+[H(x_1(t))-H(x_2(t))]. \label{gp3}
\end{align}
\end{subequations}

\section{Numerical experiments}

In this section we present numerical experiments involving our proposed adaptive strategy (\ref{GP}). We integrate the set of Eqs.  (\ref{masl}),(\ref{GP}), and we specify our sender-receiver dynamical systems to be described by the R\"ossler equation, $m=3$,
\begin{equation}
F(x_i)=\left[\begin{array}{c}
    -x_{i2}-x_{i3} \\
    x_{i1} + 0.2 x_{i2} \\
    0.2+ (x_{i1}-10 )  x_{i3}
  \end{array} \right], \label{F}
\end{equation}
$i=1,2$, coupled through $h(x_i)=x_{i1}$,  $\mathcal{H}=[1,0,0]$. Fig. \ref{A1bis} shows the average synchronization  error $E$ defined as,
\begin{equation}
E=\frac{1}{(t_2-t_1)\rho} {\int_{t_1}^{t_2} { |h(x_{2}(t))-h({x}_{1}(t))|} dt}, \label{E}
\end{equation}
from integration of Eqs. (\ref{masl}) with $\gamma(t)=\gamma$, versus $\gamma$; the normalization factor $\rho=$ $<(h(x_{1}) - <h(x_{1})>)^2>^{1/2}$, with $<...>$ indicating the time average, is calculated from  dynamics of system (\ref{F}) in the synchronous state, i.e., using dynamics from Eq. (\ref{ma}).

\begin{figure}[h]
\centerline{\psfig{figure=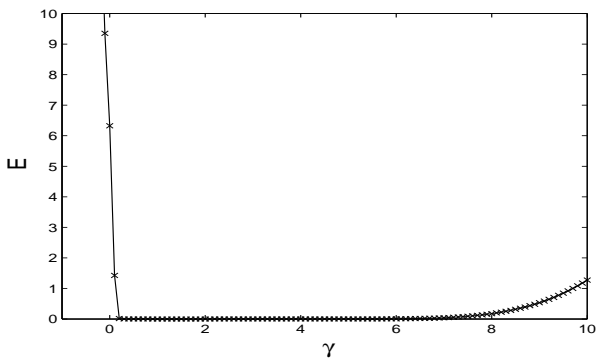,width=9cm,height=5cm}}
\caption{\small The average synchronization error $E$ from integration of Eqs. (\ref{masl}) versus (constant) $\gamma$, for $\gamma$ in the range  $[-1,10]$, $t_1=0.9 \times 10^2$, $t_2=10^2$, $\rho_1=7.45$.  \label{A1bis}}
\end{figure}

We first consider the case that $\gamma(0)$ is not far away from the stability range $\Gamma=[0.3,5]$; namely we consider two cases: in the first case we choose $\gamma(0)=0$, and in the second $\gamma(0)=5.3$. In our numerical simulations, system (\ref{ma}) and system (\ref{sl}) are evolved from random initial conditions on the R\"ossler attractor.
We choose $\nu=0.1$ so that $\nu^{-1}$ is larger than the characteristic time scale of a chaotic oscillation $T\simeq 2 \pi$, and $\beta$ to be small enough that $\gamma$ changes slowly on the time scale of the chaos, that is $\beta=10^{-4}$.

The results are shown in Fig. \ref{A2} (for the case $\gamma(0)=0$) and Fig. \ref{A3} (for the case $\gamma(0)=5.3$). In both cases, we see that for large enough $t$, $\gamma(t)$ converges to $\Gamma=[0.3,5]$, e.g., it approaches $2.32$ for the case in Fig. \ref{A2} and $1.78$ for the case in Fig. \ref{A3}. 
These values are not far away from the minimum of the master stability function  at $\gamma \simeq 1.7$ (see Fig. 1).

\begin{figure}[h]
\centerline{\psfig{figure=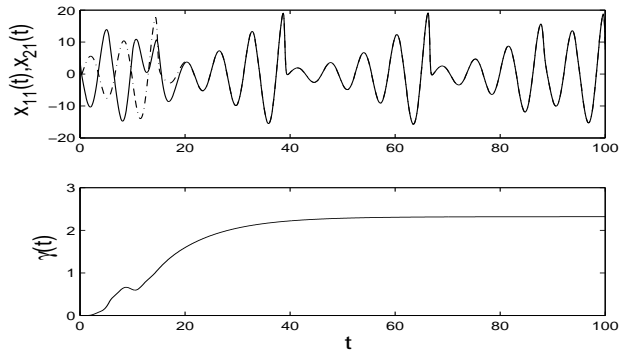,width=9cm,height=5cm}}
\caption{\small The upper plot shows the time evolution of $x_{11}(t)$ (dashed line) compared to $x_{21}(t)$ (continuous line), while the lower plot shows the time evolution of $\gamma(t)$. $\gamma(0)=0$, $\beta=10^{-4}$, $\nu=10^{-1}$.  \label{A2}}
\end{figure}

\begin{figure}[h]
\centerline{\psfig{figure=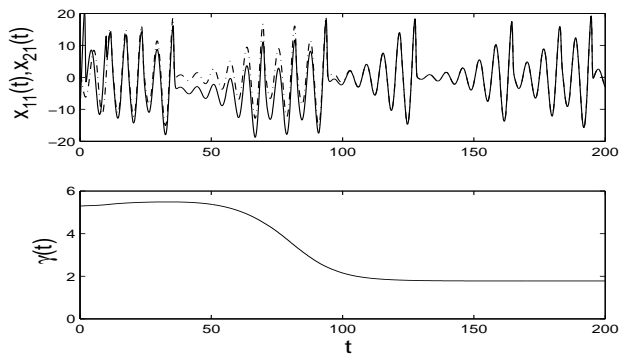,width=9cm,height=5cm}}
\caption{\small The upper plot shows the time evolution of $x_{11}(t)$ (dashed line) compared to $x_{21}(t)$ (continuous line), while the lower plot shows the time evolution of $\gamma(t)$. $\gamma(0)=5.3$, $\beta=10^{-4}$, $\nu=10^{-1}$. \label{A3}}
\end{figure}

When $\gamma(0)$ is distant from the synchronization range $\Gamma$, our proposed problem becomes more complicated. In fact, since it may take a long time for $\gamma(t)$ to reach $\Gamma$,  $x_2(t)$ may move away from $x_1(t)$ and the master stability function analysis, which describes \emph{local} stability about the synchronization manifold, may not apply anymore. 

Figure \ref{A4} shows a case in which $\gamma(t)$ is evolved from $\gamma(0)=6.5$, for $\beta=10^{-4}$ and $\nu=10^{-1}$. For this case, synchronization is not achieved, and the emergence of an unexpected phenomenon is observed (namely, the emergence of a new attractor, corresponding to $\dot{x}_{21}(t)\simeq 0$, with both $|x_{23}(t)|$ and $|x_{22}(t)|$ growing in time in such a way that $x_{23}(t) \simeq - x_{22}(t)$). In particular, from Fig. \ref{A4} we see that $\gamma(t)$ first decreases from $\gamma(0)=6.5$ to  a minimum value of about $-1.5$ and then increases again up to about $44$; it is worth noting that $\gamma(t)$ crosses twice the synchronization range $\Gamma=[0.3,5]$ and neither the first nor the second time it converges to it. We have repeated the experiment in Fig. \ref{A4} several times, eventually observing a different behavior, that is, $\gamma(t)$ increases indefinitely above $\gamma(0)=6.5$ and never goes through the synchronization range $\Gamma$ (not shown).

Based on the above observations, we conclude that, in order to improve the effectiveness of our strategy, we need to appropriately tune the rate of change of $\gamma(t)$.  In particular,
we note that (i) in the case that $\gamma(t)$ is close to $\Gamma$,  our strategy would benefit from bounding 
$\dot {\gamma}(t)$ so as to allow enough time for $\gamma(t)$ to converge in $\Gamma$;
  (ii) in the case that $\gamma(t)$ is distant from $\Gamma$, our strategy would benefit from choosing large values of $\beta$ so as to quickly adjust $\gamma(t)$ to converge towards $\Gamma$.



\begin{figure}[h]
\centerline{\psfig{figure=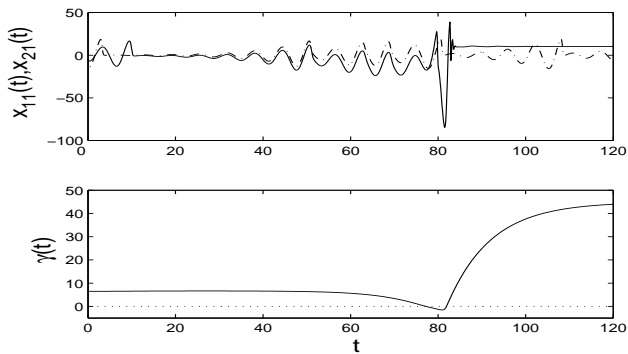,width=9cm,height=5cm}}
\caption{\small The upper plot shows the time evolution of $x_{11}(t)$ (dashed line) compared to $x_{21}(t)$ (continuous line), while the lower plot shows the time evolution of $\gamma(t)$; in the lower plot, the zero ordinate line is shown as a guide to the eye (dotted line). $\gamma(0)=6.5$, $\beta=10^{-4}$, $\nu=10^{-1}$. \label{A4}}
\end{figure}


Therefore, we propose to replace $\dot{\gamma}(t)=2 \beta G(t)$ in (\ref{gp}) by, 
\begin{equation}
\dot{\gamma}(t)=2 \beta \frac{G(t)}{1+|G(t)|}, \label{modif}
\end{equation}
where Eq. (\ref{modif}) is approximately equal to (\ref{gp}) in the case that $|G(t)| \ll 1$, while is approximately equal to $\dot{\gamma}(t)=2 \beta \times {\text{sgn}} (G(t))$ in the case that $|G(t)| \gg 1$. Note that  Eq. (\ref{modif}) corresponds to adding a saturation on $\dot{\gamma}(t)$, which now is constrained to approximately  lye in the range $[-\beta,\beta]$.

%

Figure \ref{A5} (Figure \ref{A6}) shows the results of numerical simulations in which we have tested  our modified adaptive strategy, given by Eqs. (\ref{masl}),(\ref{modif}),(\ref{gp2}),(\ref{gp3}), for a case in which $\gamma(0)=-1$ ($\gamma(0)=10$). For our experiments in Figs. \ref{A5} and \ref{A6}, we have chosen $\beta=10^{-1}$. As can be seen, in both cases $\gamma(t)$ is observed to converge to the synchronization range $\Gamma$, though it starts from initial conditions that are \emph{far away} from it.

\begin{figure}[h]
\centerline{\psfig{figure=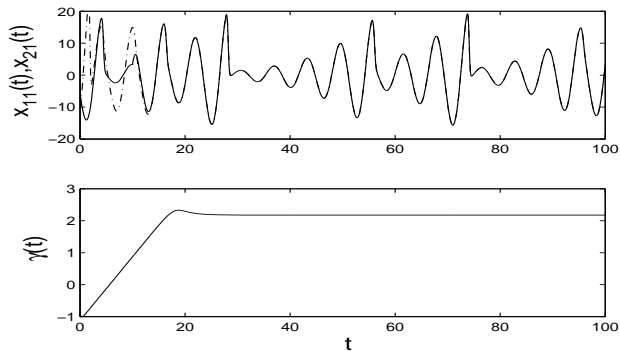,width=9cm,height=5cm}}
\caption{\small Modified adaptive strategy. The upper plot shows the time evolution of $x_{11}(t)$ (dashed line) compared to $x_{21}(t)$ (continuous line), while the lower plot shows the time evolution of $\gamma(t)$. $\gamma(0)=-1$, $\beta=10^{-1}$, $\nu=1$.  \label{A5}}
\end{figure}

\begin{figure}[h]
\centerline{\psfig{figure=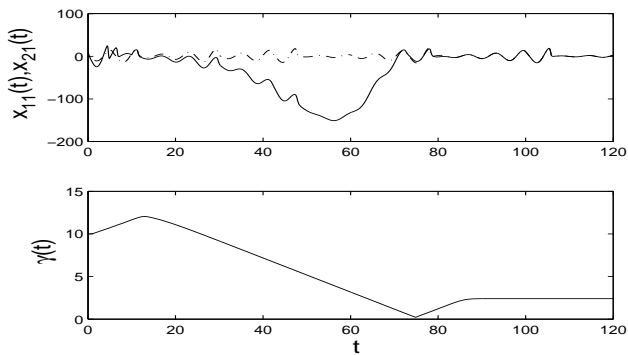,width=9cm,height=5cm}}
\caption{\small Modified adaptive strategy. The upper plot shows the time evolution of $x_{11}(t)$ (dashed line) compared to $x_{21}(t)$ (continuous line), while the lower plot shows the time evolution of $\gamma(t)$. $\gamma(0)=10$, $\beta=10^{-1}$, $\nu=1$. \label{A6}}
\end{figure}

\subsection{An experiment with the Lorenz system}

We now consider an example in which our sender-receiver dynamical systems are described by the Lorenz equation, $m=3$,
\begin{equation}
F(x_i)=\left[\begin{array}{c}
    10(x_{i2}-x_{i1}) \\
    23 x_{i1}-x_{i2}-x_{i1} x_{i3}  \\
    x_{i1} x_{i2}-x_{i3}
  \end{array} \right], \label{L}
\end{equation}
$i=1,2$. From \cite{Yang2} we know that when two Lorenz systems (\ref{L}) are coupled as in (\ref{masl}) with constant coupling $\gamma(t)=\gamma$, $h(x_i)=x_{i3}$, $\mathcal{H}=[0,0,1]$,  synchronization is stable in a bounded interval of the coupling strength $\gamma$, that is for $\gamma \in \Gamma=[1.2,6.5]$ .

Figure \ref{ALbis} shows the average synchronization  error $E$, defined in Eq. (\ref{E}), for two Lorenz systems coupled as in Eqs. (\ref{masl}), with constant coupling $\gamma(t)=\gamma$, $h(x_i)=x_{i3}$, $\mathcal{H}=[0,0,1]$, as function of the parameter $\gamma$; 
for this case, we observe that the transition to the non-synchronous state is characterized by on-off intermittency \cite{OnOff,OnOff2,Ott:Som}, which makes the use of our adaptive strategy much harder,  
since the gradient descent relation (\ref{gd}) relies on the assumption that the potential (\ref{psi}) is a monotonically increasing function of $\gamma$, for $\gamma$ moving away from $\Gamma$. However, we have implemented our adaptive strategy given by Eqs. (\ref{masl}),(\ref{modif}),(\ref{gp2}),(\ref{gp3}) and found that it can be successful applied to achieve synchronization, provided that $\gamma(0)$ is not too far away from the stability range $\Gamma=[1.2,6.5]$. Fig. \ref{A7} shows convergence of $\gamma(t)$ to values in the stability range $\Gamma$  for two different choices of the initial conditions, that is, $\gamma(0)=8$ and $\gamma(0)=0$.

\begin{figure}[h]
\centerline{\psfig{figure=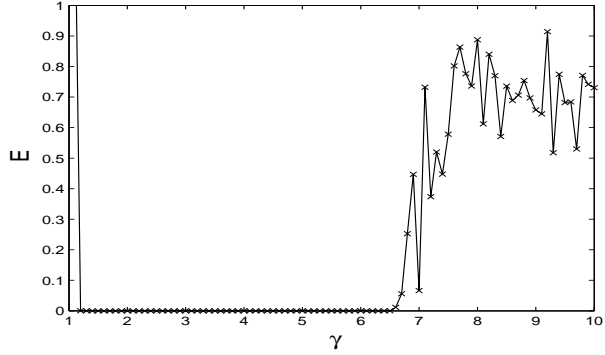,width=9cm,height=5cm}}
\caption{\small The average synchronization error $E$ obtained from integration of Eqs. (\ref{masl}) with $\gamma(t)=\gamma$, versus (constant) $\gamma$, for $\gamma$ in the range  $[0,10]$. $F(x)$ is the Lorenz equation, $h(x_i)=x_{i3}$, $\mathcal{H}=[0,0,1]$, $t_1=0.75 \times 10^3$, $t_2=10^3$.  \label{ALbis}}
\end{figure}

\begin{figure}[h]
\centerline{\psfig{figure=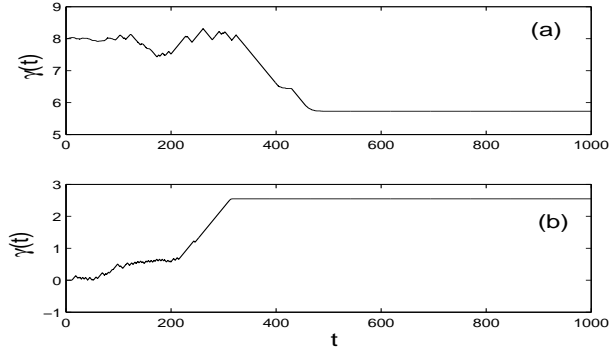,width=9cm,height=5cm}}
\caption{\small We have tested the modified adaptive strategy given by Eqs. (\ref{masl}),(\ref{modif}),(\ref{gp2}),(\ref{gp3}), for  the case of coupled Lorenz systems (\ref{L}), $h(x_i)=x_{i3}$, $\mathcal{H}=[0,0,1]$. The sender and the receiver systems are initialized from random points belonging to the Lorenz attractor. The adaptive strategy is shown to be successful in two cases, corresponding to two different choices of the initial conditions, $\gamma(0)=8$ and $\gamma(0)=0$.  Plot (a) shows the time evolution of $\gamma(t)$ from $\gamma(0)=8$, $\beta=10^{-2}$, $\nu=1$. Plot (b) shows the  time evolution of $\gamma(t)$ from $\gamma(0)=0$, $\beta=10^{-2}$, $\nu=1$. \label{A7}}
\end{figure}

\section{CONCLUSIONS}
In this paper we have considered a situation in which a sender system is unidirectionally coupled to a receiver and we have studied conditions for the receiver to adaptively evolve the strength of the coupling to achieve stable synchronization with the sender. 
We have proposed a simple adaptive strategy which has been shown to be successful in dynamically evolving the coupling strength $\gamma$ until it converges to a value in the range of stable synchronization $\Gamma$. For  cases in which the coupling strength is evolved from  initial conditions which are distant from $\Gamma$, we have proposed and numerically tested a modified adaptive strategy which includes a saturation on $|\dot{\gamma}|$.

The author is indebted to Prof. Edward Ott for insightful advices and discussions.

This work was supported by the U.S. Office of Naval Research, contract N00014-07-1-0734.


\begin{thebibliography}{17}
\expandafter\ifx\csname natexlab\endcsname\relax\def\natexlab#1{#1}\fi
\expandafter\ifx\csname bibnamefont\endcsname\relax
  \def\bibnamefont#1{#1}\fi
\expandafter\ifx\csname bibfnamefont\endcsname\relax
  \def\bibfnamefont#1{#1}\fi
\expandafter\ifx\csname citenamefont\endcsname\relax
  \def\citenamefont#1{#1}\fi
\expandafter\ifx\csname url\endcsname\relax
  \def\url#1{\texttt{#1}}\fi
\expandafter\ifx\csname urlprefix\endcsname\relax\def\urlprefix{URL }\fi
\providecommand{\bibinfo}[2]{#2}
\providecommand{\eprint}[2][]{\url{#2}}

\bibitem[{\citenamefont{Sorrentino and Ott}(2008)}]{SOTT}
\bibinfo{author}{\bibfnamefont{F.}~\bibnamefont{Sorrentino}} \bibnamefont{and}
  \bibinfo{author}{\bibfnamefont{E.}~\bibnamefont{Ott}},
  \bibinfo{journal}{Phys. Rev. Lett.} \textbf{\bibinfo{volume}{100}},
  \bibinfo{pages}{114101} (\bibinfo{year}{2008}).

\bibitem[{\citenamefont{Sorrentino and Ott}(2009{\natexlab{a}})}]{SOTT2}
\bibinfo{author}{\bibfnamefont{F.}~\bibnamefont{Sorrentino}} \bibnamefont{and}
  \bibinfo{author}{\bibfnamefont{E.}~\bibnamefont{Ott}},
  \bibinfo{journal}{Phys. Rev. E} \textbf{\bibinfo{volume}{79}},
  \bibinfo{pages}{016201} (\bibinfo{year}{2009}{\natexlab{a}}).

\bibitem[{\citenamefont{Abarbanel et~al.}(2008)\citenamefont{Abarbanel,
  Creveling, and Jeanne}}]{Abarbanel}
\bibinfo{author}{\bibfnamefont{H.~D.~I.} \bibnamefont{Abarbanel}},
  \bibinfo{author}{\bibfnamefont{D.~R.} \bibnamefont{Creveling}},
  \bibnamefont{and} \bibinfo{author}{\bibfnamefont{J.~M.}
  \bibnamefont{Jeanne}}, \bibinfo{journal}{Phys. Rev. E}
  \textbf{\bibinfo{volume}{77}}, \bibinfo{pages}{016208}
  (\bibinfo{year}{2008}).

\bibitem[{\citenamefont{Creveling et~al.}(2008)\citenamefont{Creveling, Gill,
  and Abarbanel}}]{Abarbanel2}
\bibinfo{author}{\bibfnamefont{D.~R.} \bibnamefont{Creveling}},
  \bibinfo{author}{\bibfnamefont{P.~E.} \bibnamefont{Gill}}, \bibnamefont{and}
  \bibinfo{author}{\bibfnamefont{H.~D.~I.} \bibnamefont{Abarbanel}},
  \bibinfo{journal}{Phys. Lett. A} \textbf{\bibinfo{volume}{372}},
  \bibinfo{pages}{2640} (\bibinfo{year}{2008}).

\bibitem[{\citenamefont{Quinn et~al.}(2009)\citenamefont{Quinn, Bryant,
  Creveling, Klein, and Abarbanel}}]{Abarbanel4}
\bibinfo{author}{\bibfnamefont{J.~C.} \bibnamefont{Quinn}},
  \bibinfo{author}{\bibfnamefont{P.~H.} \bibnamefont{Bryant}},
  \bibinfo{author}{\bibfnamefont{D.~R.} \bibnamefont{Creveling}},
  \bibinfo{author}{\bibfnamefont{S.~R.} \bibnamefont{Klein}}, \bibnamefont{and}
  \bibinfo{author}{\bibfnamefont{H.~D.~I.} \bibnamefont{Abarbanel}},
  \bibinfo{journal}{Phys. Rev. E} \textbf{\bibinfo{volume}{80}},
  \bibinfo{pages}{016201} (\bibinfo{year}{2009}).

\bibitem[{\citenamefont{Sorrentino and Ott}(2009{\natexlab{b}})}]{IDTOUT}
\bibinfo{author}{\bibfnamefont{F.}~\bibnamefont{Sorrentino}} \bibnamefont{and}
  \bibinfo{author}{\bibfnamefont{E.}~\bibnamefont{Ott}},
  \bibinfo{journal}{Chaos} \textbf{\bibinfo{volume}{19}},
  \bibinfo{pages}{033108} (\bibinfo{year}{2009}{\natexlab{b}}).

\bibitem[{\citenamefont{Fujisaka and Yamada}(1983)}]{FujiYama83}
\bibinfo{author}{\bibfnamefont{H.}~\bibnamefont{Fujisaka}} \bibnamefont{and}
  \bibinfo{author}{\bibfnamefont{T.}~\bibnamefont{Yamada}},
  \bibinfo{journal}{Prog. Theor. Phys.} \textbf{\bibinfo{volume}{69}},
  \bibinfo{pages}{32} (\bibinfo{year}{1983}).

\bibitem[{\citenamefont{Afraimovich et~al.}(1986)\citenamefont{Afraimovich,
  Verichev, and Rabinovich}}]{Afraim}
\bibinfo{author}{\bibfnamefont{V.~S.} \bibnamefont{Afraimovich}},
  \bibinfo{author}{\bibfnamefont{N.~N.} \bibnamefont{Verichev}},
  \bibnamefont{and} \bibinfo{author}{\bibfnamefont{M.~I.}
  \bibnamefont{Rabinovich}}, \bibinfo{journal}{Inv. VUZ Radiofiz.}
  \textbf{\bibinfo{volume}{29}}, \bibinfo{pages}{795} (\bibinfo{year}{1986}).

\bibitem[{\citenamefont{Pecora and Carroll}(1998)}]{Pe:Ca}
\bibinfo{author}{\bibfnamefont{L.}~\bibnamefont{Pecora}} \bibnamefont{and}
  \bibinfo{author}{\bibfnamefont{T.}~\bibnamefont{Carroll}},
  \bibinfo{journal}{Phys. Rev. Lett.} \textbf{\bibinfo{volume}{80}},
  \bibinfo{pages}{2109} (\bibinfo{year}{1998}).

\bibitem[{\citenamefont{Yang et~al.}(1998)\citenamefont{Yang, Hu, and
  Xiao}}]{Yang1}
\bibinfo{author}{\bibfnamefont{J.}~\bibnamefont{Yang}},
  \bibinfo{author}{\bibfnamefont{G.}~\bibnamefont{Hu}}, \bibnamefont{and}
  \bibinfo{author}{\bibfnamefont{J.}~\bibnamefont{Xiao}},
  \bibinfo{journal}{Phys. Rev. Lett.} \textbf{\bibinfo{volume}{80}},
  \bibinfo{pages}{496} (\bibinfo{year}{1998}).

\bibitem[{\citenamefont{Zhou and Kurths}(2006)}]{Zh:Ku06}
\bibinfo{author}{\bibfnamefont{C.}~\bibnamefont{Zhou}} \bibnamefont{and}
  \bibinfo{author}{\bibfnamefont{J.}~\bibnamefont{Kurths}},
  \bibinfo{journal}{Phys. Rev. Lett.} \textbf{\bibinfo{volume}{96}},
  \bibinfo{pages}{164102} (\bibinfo{year}{2006}).

\bibitem[{\citenamefont{{De Lellis}
  et~al.}(2008{\natexlab{a}})\citenamefont{{De Lellis}, di~Bernardo,
  Sorrentino, and Tierno}}]{Delellis1}
\bibinfo{author}{\bibfnamefont{P.}~\bibnamefont{{De Lellis}}},
  \bibinfo{author}{\bibfnamefont{M.}~\bibnamefont{di~Bernardo}},
  \bibinfo{author}{\bibfnamefont{F.}~\bibnamefont{Sorrentino}},
  \bibnamefont{and} \bibinfo{author}{\bibfnamefont{A.}~\bibnamefont{Tierno}},
  \bibinfo{journal}{International Journal of Computer Mathematics}
  \textbf{\bibinfo{volume}{85}}, \bibinfo{pages}{1189}
  (\bibinfo{year}{2008}{\natexlab{a}}).

\bibitem[{\citenamefont{{De Lellis}
  et~al.}(2008{\natexlab{b}})\citenamefont{{De Lellis}, di~Bernardo, and
  Garofalo}}]{Delellis2}
\bibinfo{author}{\bibfnamefont{P.}~\bibnamefont{{De Lellis}}},
  \bibinfo{author}{\bibfnamefont{M.}~\bibnamefont{di~Bernardo}},
  \bibnamefont{and} \bibinfo{author}{\bibfnamefont{F.}~\bibnamefont{Garofalo}},
  \bibinfo{journal}{Chaos} \textbf{\bibinfo{volume}{18}},
  \bibinfo{pages}{037110} (\bibinfo{year}{2008}{\natexlab{b}}).

\bibitem[{\citenamefont{Hu et~al.}(1998)\citenamefont{Hu, Yang, and
  Liu}}]{Yang2}
\bibinfo{author}{\bibfnamefont{G.}~\bibnamefont{Hu}},
  \bibinfo{author}{\bibfnamefont{J.}~\bibnamefont{Yang}}, \bibnamefont{and}
  \bibinfo{author}{\bibfnamefont{W.}~\bibnamefont{Liu}},
  \bibinfo{journal}{Phys. Rev. E} \textbf{\bibinfo{volume}{58}},
  \bibinfo{pages}{4440} (\bibinfo{year}{1998}).

\bibitem[{\citenamefont{Pikovsky}(1984)}]{OnOff}
\bibinfo{author}{\bibfnamefont{A.~S.} \bibnamefont{Pikovsky}},
  \bibinfo{journal}{Z. Phys. B} \textbf{\bibinfo{volume}{55}},
  \bibinfo{pages}{149} (\bibinfo{year}{1984}).

\bibitem[{\citenamefont{Platt et~al.}(1993)\citenamefont{Platt, Spiegel, and
  Tresser}}]{OnOff2}
\bibinfo{author}{\bibfnamefont{N.}~\bibnamefont{Platt}},
  \bibinfo{author}{\bibfnamefont{E.~A.} \bibnamefont{Spiegel}},
  \bibnamefont{and} \bibinfo{author}{\bibfnamefont{C.}~\bibnamefont{Tresser}},
  \bibinfo{journal}{Phys. Rev. Lett.} \textbf{\bibinfo{volume}{70}},
  \bibinfo{pages}{279} (\bibinfo{year}{1993}).

\bibitem[{\citenamefont{Ott and Sommerer}(1994)}]{Ott:Som}
\bibinfo{author}{\bibfnamefont{E.}~\bibnamefont{Ott}} \bibnamefont{and}
  \bibinfo{author}{\bibfnamefont{J.~C.} \bibnamefont{Sommerer}},
  \bibinfo{journal}{Phys. Lett. A} \textbf{\bibinfo{volume}{188}},
  \bibinfo{pages}{39} (\bibinfo{year}{1994}).

\end{thebibliography}
\end{document}